# PERCEVAL: a Computer-Driven System for Experimentation on Auditory and Visual Perception

Carine André, Alain Ghio, Christian Cavé, and Bernard Teston

Laboratoire Parole et Langage - UMR 6057 CNRS - Université de Provence - Aix-en-Provence, France

E-mail: carine.andre, alain.ghio, christian.cave, bernard.teston… @lpl.univ-aix.fr

## ABSTRACT

Since perception tests are highly time-consuming, there is a need to automate as many operations as possible, such as stimulus generation, procedure control, perception testing, and data analysis. The computer-driven system we are presenting here meets these objectives. To achieve large flexibility, the tests are controlled by scripts. The system's core software resembles that of a lexical-syntactic analyzer, which reads and interprets script files sent to it. The execution sequence (trial) is modified in accordance with the commands and data received. This type of operation provides a great deal of flexibility and supports a wide variety of tests such as auditory-lexical decision making, phoneme monitoring, gating, phonetic categorization, word identification, voice quality, etc. To achieve good performance, we were careful about timing accuracy, which is the greatest problem in computerized perception tests.

## 1. INTRODUCTION

Perception tests are time-consuming, not only because several speakers and listeners must be used to compensate for individual variability, but also because measurements must be repeated many times to allow for statistical validation of the results. Hence, it is necessary to computerize in auditory- and visual-perception experiments.

Different systems are available: PsyScope [1], EXPE[2], DMDX[3], Inquisit, etc. Generally, such software supports a wide variety of psychological data-collection methods. Some are easy to use, especially if they provide a graphic interface for experimentation. However, such tools are often designed for specific kinds of experiments derived from experimental psychology, and they are not suitable for other fields, particularly not for the phonetic sciences. In addition, these programs often have to be used with obsolete operating systems or special-purpose hardware (response box/sound board), which are always sources of trouble.

To develop the new version of our experimental package called PERCEVAL (French acronym for PERCeption and EVALuation), we set the following constraints:
1. Ability to simultaneously present visual and audio stimuli.
2. Response recorded through a standard keyboard or a response button box.
3. Response time measured with a temporal resolution of about 1 ms.
4. Maximum flexibility for designing perception tests.
5. Easy-to-use software with a user-friendly feature control panel.
6. Compatibility with the collective workstation [4], on which the system tests up to eight subjects simultaneously.
7. Up-to-date operating system.

## 2. SYSTEM CONSIDERATIONS

The basic system has the following components:

- A standard PC-type computer running under Windows 98/2000 OS. It is equipped with a SVGA graphic screen and a multimedia sound board.
- A custom-made button box can be added to record the subject's responses. These boxes are built with the printed circuit board (PCB) of a USB joystick or game pad. PCBs are inserted in plastic cases with up to 8 push buttons. These buttons are short stroke, sharp contact, fast action microswitches specially suited to measuring reaction times.
- For auditory stimulation, we generally use an audio amplifier and Beyer BT100 electrodynamic earphones. They were chosen for their durability, high sensitivity, and response curves [4].

The system can be used on a portable PC computer.

## 3. TIMING ACCURACY

Timing accuracy is important in perception tests. It can be critical in priming experiments in psychology, for example, where the delay between two views must be precisely controlled. Response time is also important. It can inform us about the difficulty of the task, and it may be the only information when the effect is saturated [5].

### 3.1 Why is timing accuracy difficult to control on computer-driven perception tests?

Controlling and measuring time on computer-driven perception tests has been studied for a long time [1,2,3,4,6,7]. Programs that run on a single-task operating system (e.g. DOS) have the advantage of being able to perfectly time of the computer-driven test. In fact, in such an environment, the program runs sequentially without "long interruptions" and most of the computer's computational power is dedicated to the test program. On the other hand, with multitask operating systems (e.g. Microsoft Windows), a program may be interrupted for a long time because the computer shares its computational power among all applications running at the same time. In computer-driven perception tests, this phenomenon can be catastrophic if, for example, the dedicated program is inactive when the screen has to be refreshed or when the subject is responding. This can cause timing inaccuracy.

*3.2 Why is it necessary to develop packages on multitask operating systems like Windows?*

Programs developed for DOS generally require dedicated hardware [2]. Most displays or playbacks are written directly to a specific video or sound card to achieve the performance required of experimental software. Moreover, new video and sound cards are being released without consideration of DOS support, and more generally, recent PC computers no longer support DOS, which is quickly disappearing. It is clear that for a new application to have an acceptable life span, it has to be based on an up-to-date environment like Windows [3].

*3.3 Why are some solutions unable to obtain timing accuracy?*

One way to limit timing inaccuracy on multitask operating systems is to give a high priority level to the dedicated program, which means that it is executed frequently in the computer task list. But even with this precaution, there is no guarantee that it will not be interrupted for a critical duration by another high-priority task (e.g. drivers managing hardware). The use of "timers" may be a solution for controlling timing accuracy. A timer is a way to execute a function regularly. For example, it can be used to detect whether the subject has responded (by systematic scanning of a response device). But standard timers running on Microsoft Windows have a resolution of about 50 ms, which is very poor for precisely measuring response time. Multimedia timers offer a better resolution (about 1 ms) but this precision is not guaranteed.

*3.4 Why does command latency vary?*

Another problem is related to the latency of some commands. The latency is the delay between the moment when a command is launched (e.g. changing a view, playing a sound) and the moment when the command is actually executed. For example, on a Pentium III 1.2 GHz, the C command *PlaySound("stimulus.wav")* to play back a wave file has a latency of 100 ms. This means that if a timer that measures response time is started just before the *PlaySound* instruction, it will have an error of about 100 ms because the actual audio stimulus started 100 ms later than planned. One solution could be to measure all the critical latencies once and for all, and take them into account in the results. But the problem is more complicated. In fact, these latencies can change, depending on a large set of factors, which means that this solution is inadequate.

*3.5 How can timing accuracy be obtained?*

First of all, the computer in charge of the perception test must be dedicated to this job only during the test [3]. The operator must be sure to remove all tasks that can use up processing time (e.g. anti-virus, network, CD-Rom). But this is still not enough.

To obtain timing accuracy on the Perceval workstation, we use a well-suited technology called DirectX. DirectX provides a set of programming interfaces for designing high-speed applications while obtaining the benefits of direct access to the hardware. This software development kit includes several components. DirectSound provides low-latency mixing, hardware acceleration, and direct access to the sound device. DirectDraw allows one to directly manipulate display memory, hardware blitter, and the flipping surface support. These two components provide the capability of reducing the latency of playback and display functions at a high ratio because they are close to the hardware. Moreover, it is possible to prepare a set of time-consuming tasks before the real stimulation. For example, as a first step (preliminary), one can load a wave file or an image file in memory and then, as a second step (the real test), play back the sound and flip the screen image very rapidly. For comparison (see section 3.4), on a Pentium III 1.2 GHz, the DirectSound function "Play" has a latency of 0.5 ms. The third component used is DirectInput, which enables an application to gain access to input devices, including the mouse, keyboard, and joystick, even if the application is in the background. This capability can be implemented by using buffered data, which is a record of events that are stored until the application retrieves them.

## 4. MODULAR ARCHITECTURE

The Perceval package includes several specialized applications: stimulus creation tools, experiment design wizard, subject manager, test module, scoring module, etc.

Tests are controlled by scripts, a technique already used in the S.O.A.P. system [7]. In fact, the system's core software resembles that of a lexical-syntactic analyzer, which reads and interprets the script files sent to it. The execution sequence (trial) is modified according to the commands and data received. This type of operation provides a great deal of flexibility and supports a wide variety of tests. A script is a text file composed of different sections: information, data, trial events, and settings (Figure 1). These files can be edited manually by an advanced user who knows the syntax. A novice can also follow the instructions given by an application design wizard with a menu-driven interface.

For each subject, a response file is generated containing the necessary information for processing the data. These files are written in table format to make it easy to export them to spreadsheets (Figure 2) or statistical analysis packages.

## 5. EXPERIMENTAL DESIGN AND RESULTS

*5.1 Script example adapted to minimal pairs (Figure 1)*

This test consists in playing the sounds, displaying word pairs, and recording the subject's responses. The results can be improved by including the features and contexts of the consonants tested.
Usually, a script is divided into four major sections:

```
[INFORMATION]
AUTHOR=A. Ghio & C. André
DATE=14/01/2003
TITLE=Paires Minimales réduites
VERSION=3.0.2.0

[TRIAL_DATA]
TRIAL1=<1) main 2)bain>          <bain.wav>        <Choix2>        <-nasal>           <E~>
TRIAL2=<1)bain 2)main>           <main.wav>        <Choix2>        <+nasal>           <E~>
TRIAL3=<1)bal 2)val>             <bal.wav>         <Choix1>        <+interrompu>      <aa>
TRIAL4=<1)val 2)bal>             <val.wav>         <Choix1>        <-interrompu>      <aa>
...          #1                      #2                #3               #4             #5

[TRIAL_EVENTS]
X10=BEGIN
X20=DISPLAY_TEXT<#1>
X30=PLAY_SOUND<#2>
X40=GET_INPUT<DELAY 2000>
X50=END

[SETTINGS_GROUP1]
INSTRUCTION_FORMAT=<Pairemin.txt>                *1
TRAINING_ORDER=<1 3 4 6>                         *2
TRIAL_ORDER=<RANDOM>                             *3
TEXT_FORMAT=<FONT Arial><SIZE 30><BKCOLOR 0x0000FF><TXTCOLOR 0xFFFF00><POSITION HCenter|VCenter>   *4
INPUT=<Choix1 CK_1 VK_NUMPAD1 BK_01><Choix2 CK_2 VK_NUMPAD2 BK_02>  *5
CORRECT=<#3>     *6
PAUSE=0          *7
RESPONSE_FORMAT=<$SUBJECT><$TRIAL><#1><#2><#3><$RESPONSE><$ERROR><#4><#5><$RTIME>  *8
```

**Figure 1**. Script example

The *[INFORMATION]* section contains general information: the name of the author, the creation date, the version used, the title given.

The *[TRIAL_DATA]* section defines all data one needs during the experiment for each trial executed. Each line of data is divided into columns.

The *[TRIAL_EVENTS]* section defines the event sequence for each trial played. The first order in this section is always *BEGIN* and last is always *END*.

The command on line X20 displays the contents of the first column (#1) of the trial data on the screen. In the above example, if the trial played is TRIAL1, then *1)main 2)bain* is displayed.

The command on line X30 plays the wave file located in the second column (#2) of the trial data. In our example, if the trial played is TRIAL1, the sound *bain.wav* is played.

The command on line X40 starts the recording of the response. In this example, two seconds are left for the subject to respond.

The *[SETTINGS_GROUP1]* section defines the experiment configuration (display, trial order, input, pause, etc.). Several groups can be included in the same script.

At the beginning of the experiment, instructions are displayed on the screen. The instructions are contained in a text file (*1).

A training phase can be executed. The order of the trials to be run is defined in *2.

During the actual test phase, the order of the trials can be fixed or randomized (*3).

The format (font, size, etc.) of the text displayed on screen is set in *4.

The authorized keys for responding are listed in *5: a standard keyboard or a button box can be used for the test.

Depending on the type of test, the concept of correct/incorrect response can be introduced (*6). In our example, this information is in the third column (#3) of the trial data.

The pause value (*7) defines the time between trials.

The results written in the response file are formatted in *8. Figure 2 shows an example of a response file obtained with the previous script (Figure 1).

| A | B | C | D | E | F | G | H | I | J |
|---|---|---|---|---|---|---|---|---|---|
| $$ | $T | #1 | #2 | #3 | $RESP | $ER | #4 | #5 | $RT |
| ca | 4 | 1)val 2)bal | val.wav | Choix1 | Choix2 | err | -interrompu | aa | 1072 |
| ca | 1 | 1)main 2)bain | bain.wav | Choix2 | Choix2 | ok | -nasal | E~ | 748 |
| ca | 14 | 1)latte 2)ratte | ratte.wav | Choix2 | Choix1 | err | +compact | aa | 1012 |
| ca | 17 | 1)veste 2)ouest | veste.wav | Choix1 | Choix1 | ok | -vocalique | EE | 830 |
| ca | 20 | 1)furent 2)surent | surent.wav | Choix2 | Choix2 | ok | +grave | yy | 1066 |
| ca | 12 | 1)dette 2)bette | bette.wav | Choix2 | Choix2 | ok | -grave | EE | 1218 |
| ca | 8 | 1)bosse 2)gosse | bosse.wav | Choix1 | xxx | xxx | +compact | oo | 0 |

**Figure 2**. Response file for minimal pairs (Excel file)

A: $SUBJECT specifies the subject code.
B: $TRIAL defines the trial number.
C: #1 is the contents of the first column of the trial data for that trial. In this example, it is the text displayed.
D: #2 is the contents of the second column of the trial data for that trial. In this example, it is the sound played.
E: #3 is the contents of the third column of the trial data for that trial. In this example, it is the correct response.
F: $RESPONSE is the response given by the subject.
G: $ERROR specifies whether the subject's response is correct or incorrect ("ok" if it correct, "err" if not).
H: #4 is the contents of the fourth column of the trial data for that trial. In this example, it is the consonant feature.
I: #5 is the contents of the fifth column of the trial data for that trial. In this example, it is the vocalic context of the consonant.
J: $RTIME is the subject's reaction time.

*5.2 Example of a script for varying the volume of a sound (Figure 3)*

The subject must determine whether the sound heard (aaa.wav) seems strong or weak (Fort/Faible). We vary the volume of the sound using the *VOLUME* order. In the example shown in Figure 3, if the trial executed is the first one, the sound is played as it was recorded. If the trial is the second one, the sound is played at 3 db less than it was recorded.

```
[TRIAL_DATA]
TRIAL1=<0>
TRIAL2=<-3>
TRIAL3=<-6>

[TRIAL_EVENTS]
X10=BEGIN
X20=DISPLAY_TEXT <1)Fort    2)Faible>
X30=PLAY_SOUND <aaa.wav><VOLUME #1>
X40=GET_INPUT
X50=END
```

**Figure 3**. Variation of a sound's volume

*5.3 Example of a script using gating (Figure 4)*

The subject listens to one of the two words ("bêle" or "bête") for a variable amount of time (gating). The beginning (*TIME_BEGIN*) and the end (*TIME_END*) of the sound listening time are defined. Here, the sound is played between the beginning and ending times defined by the third parameter in the trial data list.

```
[TRIAL_DATA]
TRIAL1=<bêle><bele.wav><200>
TRIAL2=<bêle><bele.wav><250>
TRIAL3=<bêle><bele.wav><275>
TRIAL4=<bête><bette.wav><200>
TRIAL5=<bête><bette.wav><250>

[TRIAL_EVENTS]
X10=BEGIN
X20=DISPLAY_TEXT<1) bêle    2) bête>
X30=PLAY_SOUND<#2><TIME_BEGIN 0><TIME_END #3>
X40=GET_INPUT<DELAY 2000>
X50=END
```

**Figure 4**. Use of gating

*5.4 Example of a script using feedback (Figure 5)*

The subject sees images with an associated word, and must decide whether the word corresponds to the image. A sound is emitted to tell the subject whether the answer given was right or wrong (sound feedback).

```
[TRIAL_DATA]
TRIAL1=<catre><faute><catre.bmp>
TRIAL2=<glace><faute><glace.bmp>
TRIAL3=<horloche><faute><horloche.bmp>
TRIAL4=<sourus><faute><sourus.bmp>

[TRIAL_EVENTS]
X10=BEGIN
X20=DISPLAY_FILEBMP<#3>
X40=GET_INPUT<DELAY 3000>
X50=END

[SETTINGS_GROUP1]
...
SOUND_FEEDBACK=<POSITIVE clap.wav><NEGATIVE glass.wav>
```

**Figure 5**. Use of feedback

# 6. CONCLUSION

The computer-driven speech assessment system we developed is used to evaluate auditory-lexical decisions, phoneme monitoring, gating, phonetic categorization, word identification, voice quality, speech intelligibility, and so on. It was designed to automate the various operations involved in setting up experiments, presenting stimuli, and recording subject responses on a single PC computer. The experience acquired during the development process will be applied to enhancing the performance and reliability of our multiple-listener workstation, which can test as many as eight subjects at a time.

*PERCEVAL is available on the website of the "Speech and Language Laboratory" at the following address: www.lpl.univ-aix.fr*